\documentclass[aps,prb,twocolumn,showpacs]{revtex4}

\usepackage{amsmath}
\usepackage{graphicx}
\usepackage{setspace}
\usepackage{bm}
\usepackage{dcolumn}
\usepackage{ulem}

\begin{document}

\title{Mathematical model of vortex penetration phenomenon}

\author{Rongchao Ma} \email{marongchao@yahoo.com}
\affiliation{Department of Physics, University of Alberta, Edmonton, AB, Canada T6G 2G7}

\date{\today}

\begin{abstract}
Vortex penetration affects the stability of a superconducting system and limits the possible application of the system. However, the mathematical description to this phenomenon is currently unavailable. Here I presented a mathematical model in which I considered the effects of bulk pinning and internal field repulsive force on vortex hopping. Thereafter, I proposed a series expansion to the activation energy and derived a general formula to describe the time dependence of the vortex penetration process. With these formulas, I can analyze the experimental data and calculate the activation energy of the vortex penetration phenomenon. The results are accurate for the time dependence of the internal field measurements in a $Bi_2Sr_2CaCu_2O_{8+x}$ superconductor.
\end{abstract}


\maketitle

\section{Introduction}

An applied magnetic field can penetrate into a type-II superconductor in the form of quantized vortices (each carry a single flux quantum $\Phi_0$) when the applied magnetic field increases to a value above the lower critical field of the superconductor.\cite{Abrikosov,Bascom,Kleiner,Gammel1,Murray,Goa} The vortices first penetrate into the superconductor from the surfaces which are parallel to the applied magnetic field, and then go to the center of the superconductor.\cite{Bean,Blois} Vortex motion is usually retarded \cite{Joseph} during the penetration process because the vortices are subjected to surface attractive force \cite{Bean}, pinning forces, internal field repulsive force and damping force \cite{Bardeen,Gurevich}. The internal magnetic field (or penetrated magnetic field) increases gradually and finally reaches a saturated value. De Gennes \cite{de Gennes} first studied the vortex penetration in terms of free energy. A number of works studied the vortex entry conditions \cite{Kramer,Kato,Bolech,Berdiyorov,Baelus1,Baelus2,Hernandez,Sela}, and other works investigated the vortex penetration into the bulk of superconductors over the Bean-Livingston surface barrier \cite{Burlachkov,Clem,Mawatari,Elistratov,Pissas,Wang,Erdin}. The experimental studies of the vortex penetration were generally carried out with Hall sensors \cite{Pissas} and magneto-optical imaging technique \cite{Goa,Kasahara} which provides direct observations of vortex front.

It is known that vortex entry conditions and vortex penetration over surface barrier were studied. However, the Bean-Livingston surface barrier reduces exponentially with respect to the distance to the surface of the superconductor.\cite{Bean} The vortex motion inside a macro size superconductor is then mainly determined by the bulk pinning and internal field repulsive force. The theoretical description to this phenomenon is currently unavailable. On the other hand, the recent experimental observations \cite{Ma1} have shown that the internal magnetic field of a high-$T_c$ superconductor is strongly time dependent after a magnetic field (below irreversible field) is applied to the superconductor. The time scale depends on the sample geometry, temperature and applied field magnitude. This time dependence implies that a vortex system is unstable during the vortex penetration process and a superconducting device may not work properly at this stage. For the purpose of application, therefore, it is important to construct a mathematical model which includes the effects of the bulk pinning and internal field repulsive force, and can predicts the time scale of the vortex penetration process.

In this article I theoretically show that the vortex penetration process is time dependent because of the surface attractive force, bulk pinning force and internal field repulsive force. First, I discussed the field dependence of activation energy of the vortex penetration process. Next, I discussed the time dependence of activation energy based on the Arrenhius relation. Finally, I derived a time dependent equation for the internal field in a vortex penetration process.

\section{Model}

As mentioned before, the vortices inside a type-II superconductor are subjected to various forces and vortex motion is generally retarded. The experimental observations \cite{Ma1} have shown that, at lower temperatures and under small applied fields, the time scale of vortex penetration process in a superconducting crystal sample extends to infinity long. Therefore, it is reasonable to assume that the vortices are at some metastable states under an intermediate applied field $H_{c1}< H_a < H_{irr}$, where $H_{c1}$ is lower critical field and $H_{irr}$ is irreversible field. In this sense, the vortex penetration process can be regarded as a vortices diffusion process, i.e., a process of vortices hopping between adjacent pinning centers. In an analogy to that of flux relaxation \cite{Anderson}, we can described the vortex penetration process with the Arrhenius relation: $\nu=\nu_0 e^{-U_a/kT}$, where $\nu_0$ is attempting frequency, $U_a$ is the activation energy of the vortices, $k$ is the Boltzmann constant and $T$ is temperature.\cite{Tinkham} This relation implies that the activation energy $U_a$ plays an important role in the vortices hopping.

In this work I am going to use some similar concepts and mathematical tools used in the studies of flux relaxation. But it should be emphasized that the physical contents of these two processes are completely different. In vortex penetration process the vortices are pushed into the superconductor by external driving force; however, in flux relaxation process the vortices jump out of the superconductor because of various reasons \cite{Ma2}. Let us first discuss the possible expression of the field dependence of activation energy $U_a$ in the vortex penetration process.

\subsection{Field dependence of activation energy}

According to Bean and Livingston \cite{Bean}, a vortex close to the surface of a superconductor is subjected to attractive imaging force and external field repulsive force. The addition of these two forces results in the surface barrier to the vortex penetration, $U_{BL}$, which is determined by the sample geometry and external field magnitude $H_a$. Early studies have shown that \cite{Bean}
\begin{equation}
\label{Us}
U_{BL}(x) = \left(\frac{\Phi_0}{4 \pi \lambda}\right)^2 K_0\left(\frac{2x}{\lambda}\right) - \frac{\Phi_0}{4\pi} H_a e^{-x/\lambda} 
\end{equation}
where $\Phi_0$ is flux quantum, $\lambda$ is penetration depth and $x$ is the distance between the vortex and sample surface. The first term on the right side of Eq.(\ref{Us}) is caused by the attractive imaging force, and the second term on the right side is caused by the applied field driving force. Eq.(\ref{Us}) shows that $U_{BL}(x)$ reduces exponentially with respect to $x$, that is, $ x \rightarrow \infty $, $U_{BL}(x) \rightarrow 0 $. Thus, the vortices away from the sample surface are almost uninfluenced by the surface barrier.

On the other hand, a vortex inside a superconductor is subjected to bulk pinning force and internal field repulsive force, which prevents the vortices motion and reduces the vortices hopping rate. According to the Arrhenius relation, this internal field repulsive force increases the activation energy. Therefore, the activation energy is an increasing function of the internal field $B$. Let $U_b$ be the activation energy related to the bulk pinning and internal field repulsive force, we have $dU_b/dB > 0$.

The exact form of $U_b(B)$ is usually unknown because of the complexity of the interaction between vortices and pinning centers, as well as the reaction of vortices to external field driving forces. But from a pure mathematical consideration, we can express $U_b(B)$ as a Taylor series of $B$ (or magnetization $M$).\cite{Ma2} Let us now write out the series expression of $U_b(B)$ explicitly,
\begin{equation}
\label{Ub}
U_b(B) = U_c + \sum\limits_{l=1}^n a_l B^l
\end{equation}
where $U_c=U_b(0)$, $ a_1=U'_b(0)$, $a_2=U''_b(0)/2!$, $\cdots$, $a_n=U^{(n)}_b(0)/n!$. The parameter $U_c$ is the pinning potential inside the bulk of the superconductor. The coefficients $a_l$ ($l>2$) represent the weight of the contribution from inelastic deformation and interaction between the vortices, $a_2$ represents the weight of the contribution from elastic deformation of the vortices and $a_1$ represents the weight of the contribution from Lorentz force.\cite{Ma2} These coefficients are functions of temperature $T$ and can be denoted as $U_c(T, \lambda, \xi)$ and $a_l(T, \lambda, \xi)$.

The total activation energy of a vortex, $U_a(B)$, is the summation of $U_{BL}$ and $U_b(B)$, that is
\begin{equation}
\label{UaField}
U_a(B) = U_{BL} + U_b(B) = U_0 + \sum\limits_{l=1}^n a_l B^l
\end{equation}
where $U_0=U_{BL}+U_c$ is the activation energy at vanishing internal field, or the activation energy at time $t=0$. $U_0$ is determined by the sample geometry, external field and pinning ability. 

From Eq.(\ref{Us}) we see that the activation energy $U_a(B)$ is a local function, i.e., a function of position $x$. This is not explicitly shown in Eq.(\ref{UaField}) because we are interested in the field dependence of $U_a$ in the current work.

\subsection{Time dependence of activation energy}

The above discussion has shown that in vortex penetration process the activation energy $U_a$ is an increasing function of the internal field $B$. It is also known that $B$ increases gradually with time $t$. Therefore, $U_a$ is an increasing function of $t$.

The rate of change of the internal field, $dB/dt$, is proportional to vortex hopping rate. On the other hand, $B$ is an increasing function of $t$ in the vortex penetration process, i.e., $dB/dt>0$ (in relaxation $dB/dt<0$). According to the Arrhenius relation, we have \cite{Geshkenbein}
\begin{equation}
\label{UaDiff}
\frac{dB}{dt} = C e^{-U_a/kT}
\end{equation} 
where $C$ is a positive proportional constant. Eq.(\ref{UaDiff}) can also be derived from the diffusion equation of flux flow (conservation of flux).\cite{Beasley} Comparing to that of flux relaxation \cite{Geshkenbein}, Eq.(\ref{UaDiff}) has differences in sign and initial conditions. These cause the differences between the mathematical models of vortex penetration and flux relaxation.

The exact solution of Eq.(\ref{UaDiff}) is currently unavailable, but we can find an approximate solution to it. Consider applying a magnetic field to a superconductor at time $t=0$. The activation energy of a vortex at this moment is $U_0$, which is also the activation energy at zero internal field (See Eq.(\ref{UaField})). As time increases to $t$, the activation energy increases to $U_a$. Rewrite Eq.(\ref{UaDiff}) and integrate it on both sides,

\begin{equation}
\label{UaDiff2}
\int_{U_0}^{U_a} e^{U_a/kT}dU_a = \int_0^t C\frac{dU_a}{dB} dt
\end{equation}

With logarithmic accuracy \cite{Geshkenbein}, we obtain the following equation:
\begin{equation}
\label{UaTime}
U_a(t) = U_0 + kT ln\left(1 + \frac{t}{\tau e^{U_0/kT}} \right)
\end{equation} 
where $\tau = kT/[C(dU_a/dB)]$ is a short time scale parameter \cite{Feigel'man2}. Eq.(\ref{UaTime}) shows that $U_0$ has a strong effect on the time dependence of the activation energy $U_a(t)$.

\subsection{Time dependence of internal field}

Our final purpose is to obtain the time dependence of internal field $B(t)$. Since the vortex motion in a vortex penetration process is retarded by various forces, the internal field increases gradually with increasing time. This relation can be obtained by canceling out the activation energy $U_a$ from Eq.(\ref{UaField}) (field dependence of $U_a$) and Eq.(\ref{UaTime}) (time dependence of $U_a$), that is,
\begin{equation}
\label{WvsJ}
w(t)= \sum\limits_{l=1}^n a_l B^l(t)
\end{equation}
where
\begin{equation}
\label{FunctionW}
w(t)= kT ln\left(1 + \frac{t}{\tau e^{U_0/kT}} \right)
\end{equation}

To find out the time dependence of internal field $B(t)$, let us now invert Eq.(\ref{WvsJ}) by expanding $B[w(t)]$ as a series of $w(t)$,
\begin{equation}
\label{BvsW}
\begin{aligned}
B[w(t)] = \sum\limits_{l=1}^n b_l w^l(t) = \sum\limits_{l=1}^n b_l \left[kT ln\left(1 + \frac{t}{\tau e^{U_0/kT}} \right) \right]^l
\end{aligned}
\end{equation}

Similar to earlier studies \cite{Ma2}, we have the coefficients $b_l$'s \\
\begin{widetext}
\begin{equation}
\label{bns}
\begin{aligned}
&b_1=\frac{1}{a_1} \\
&b_2=\frac{1}{a_1^2}\left(-\frac{a_2}{a_1}\right) \\
&b_3=\frac{1}{a_1^3}\left[2 \left(\frac{a_2}{a_1}\right)^2 - \left(\frac{a_3}{a_1}\right)\right] \\
   &\cdots \\
&b_l=\frac{1}{a_1^l} \frac{1}{l} \sum\limits_{s,t,u \cdots} (-1)^{s+t+u+\cdots} \cdot \frac{l(l+1)\cdots(l-1+s+t+u+\cdots)}{s!t!u!\cdots} \cdot \left(\frac{a_2}{a_1}\right)^s \left(\frac{a_3}{a_1}\right)^t \left(\frac{a_4}{a_1}\right)^u \cdots \\ 
&\cdots \\   
\end{aligned}
\end{equation}
\end{widetext}
where $s+2t+3u+\cdots=l-1$. On considering the symmetry between Eq.(\ref{WvsJ}) and Eq.(\ref{BvsW}), we obtain the coefficients $a_l$ by doing a commutation to the coefficients $b_l \leftrightarrow a_l$, that is 

\begin{widetext}
\begin{equation}
\label{ans}
a_l= \frac{1}{b_1^l} \frac{1}{l}\sum\limits_{s,t,u \cdots} (-1)^{s+t+u+\cdots} \cdot \frac{l(l+1)\cdots(l-1+s+t+u+\cdots)}{s!t!u!\cdots} \cdot \left(\frac{b_2}{b_1}\right)^s \left(\frac{b_3}{b_1}\right)^t \left(\frac{b_4}{b_1}\right)^u \cdots \\
\end{equation}
\end{widetext}

Eq.(\ref{BvsW}) describes the time dependence of internal field $B(t)$. This relation can also be measured experimentally.\cite{Ma1} With these relations we can calculate the field dependent activation energy $U_a(B)$ using the following procedure: first, fitting the experimental data with Eq.(\ref{BvsW}), we obtain the fitting parameters $U_0$ and $b_l$ as shown in Figure 1. Next, substituting $b_l$ into Eq.(\ref{ans}), we obtain the coefficients $a_l$. Finally, substituting $U_0$ and $a_l$ into Eq.(\ref{UaField}), we obtain the activation energy $U_a(B)$.

The activation energy $U_a(B)$ is a combined response to the internal field $B$. It includes the contributions from the surface barrier, deformation of vortices, interaction between vortices and possibly other unknown sources. Figure 1 shows that Eq.(\ref{BvsW}) is accurate for the experimental data of a $Bi_2Sr_2CaCu_2O_{8+x}$ single crystal.

\begin{figure}[htb]
\begin{center}
\includegraphics[width=70mm, height=60mm]{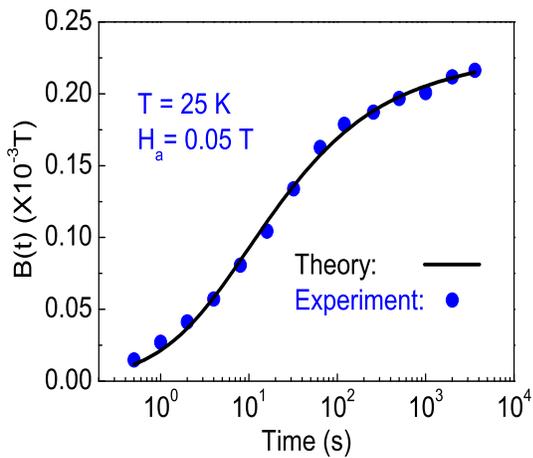}
\caption{\label{figure1} (Color online) Time dependence of vortex penetration. The scattering points are the experimental data of a $Bi_2Sr_2CaCu_2O_{8+x}$ single crystal, at 25 K under an applied magnetic field of 0.05 T. (The data is from Ref.~\onlinecite{Ma1}) The solid black line is the theoretical fit. The fitting results are:
$B(t)=b_1w(t)+b_2w^2(t)+b_3w^3(t)$, where $w(t)= (25k) \times ln \{1+t/[ \tau e^{U_0/(25k)} ]\}$, $U_0=(1.27 \pm 0.01)\times(25k)$, $\tau = 0.82\pm0.01$, $b_1=(0.75\pm0.01)\times10^{-4}/(25k)$, $b_2=(-0.09\pm0.01)\times10^{-4}/(25k)^2$, $b_3=0.00$.}
\end{center}
\end{figure}

At lower temperatures and lower applied fields, the inelastic deformation and interaction between vortices is not significant. Therefore, we can simplify the activation energy $U_a(B)$ in a manner analogous to that used in the studies of flux relaxation \cite{Ma2}.

First, let us consider putting $a_l=0$ ($l>2$) in Eq.(\ref{UaField}). This is equal to ignoring the inelastic deformation and interaction between vortices. Thus, we obtain the non-interacting elastic vortices which are described by the quadratic activation energy
\begin{equation}
\label{UaQuadratic}
U_a(B) = U_0 + a_1 B + a_2 B^2
\end{equation}

Substituting Eq.(\ref{UaTime}) into Eq.(\ref{UaQuadratic}), we have (choose one of the solutions which is an increasing function of time),
\begin{equation}
\label{jElastic}
B(t) = \frac{a_1}{2a_2} \left[ \sqrt{1 + 4\frac{a_2}{a_1^2} w(t)} - 1 \right]
\end{equation}
where $w(t)$ is defined by Eq.(\ref{FunctionW}).

Further putting $a_2=0$ in Eq.(\ref{UaQuadratic}) is equal to assuming that the vortices have very large elastic modulus and the elastic deformation can be ignored. We obtain the non-interacting rigid vortices which are described by the linear activation energy
\begin{equation}
\label{UaLinear}
U_a(B) = U_0 + a_1 B
\end{equation}

Substituting Eq.(\ref{UaTime}) into Eq.(\ref{UaLinear}), we have 
\begin{equation}
\label{jLinear}
B(t) = \frac{1}{a_1} w(t)
\end{equation}

\section{Discussion}

1. \textit{Measurement of $U_c$ and $U_{BL}$}.---
The bulk pinning potential $U_c$ can be obtained by measuring the time evolution of internal field $B(t)$ at a position $x >> \lambda$ inside the sample where $U_{BL}(x)=0$. Fitting the experimental data $B(t)$ with Eq.(\ref{BvsW}), we have the value of $U_c$ at the position $x$. In case of random pinning centers, $U_c$ is a constant and can be represented by the statistical average value $f \xi N^{1/2}$, where $f$ is the pinning force of individual pinning center, $\xi$ is coherence length and $N$ is the number of the pinning centers in a correlation volume.\cite{Larkin,Tinkham}. 

The surface barrier $U_{BL}$ can be calculated using Eq.(\ref{Us}), but we can also measure it with the help of Eq.(\ref{BvsW}). First, using the above stated procedure to measure the bulk pinning potential $U_c$. Next, fixing a sensor at a position $x'$ close to the sample surface and measuring $B(t)$, we obtain the value of $U_0(x')$ using Eq.(\ref{BvsW}). Finally, using relation $U_{BL}(x')=U_0(x')-U_c$, we can find out the value of $U_{BL}(x')$.

If a superconductor is anisotropic, the superconducting parameter $\lambda$ is anisotropic. According to Eq.(\ref{Us}), the surface barrier $U_{BL}$ is also anisotropic. Therefore, when estimating the surface barrier we must have a clear idea about which crystalline axis is perpendicular to the surface.

2. \textit{Inflection point of $B(t)-t$ curve}.---
As shown in Figure 1, the $B(t)-t$ curve displays a concave shape at short time and then changes to a convex shape with increasing time. Let us now discuss the possible reason for this phenomenon.

The fitting results indicate that it is accurate enough by keeping terms up to second order. Thus, we can write out the time dependent internal field as $B(t)=b_1w(t)+b_2w^2(t)$. The second derivative of $B(t)$ is, 
\begin{equation}
\label{SecDerivation}
\frac{d^2B}{dt^2}= g(t) \left[1-\frac{b_1}{2b_2kT}-ln\left(1+\frac{t}{\tau e^{U_0/kT}}\right)\right]
\end{equation}
where $g(t)= 2b_2 [kT/(\tau e^{U_0/kT}+t)]^2$. Putting $d^2B/dt^2=0$, we obtain the inflection point time 
\begin{equation}
\label{Inflection}
t^*=\tau e^{U_0/kT}[e^{1-b_1/(2b_2kT)}-1]
\end{equation}

If $t<t^*$, then $d^2B/dt^2>0$; the curve is concave. If $t>t^*$, then $d^2B/dt^2<0$; the curve is convex. One can easily see that $t^*$ is a decreasing function of temperature $T$ (note that $b_1$ and $b_2$ has different sign). On the other hand, $t^*$ is an increasing function of $U_0$ and $U_0$ is a decreasing function of the applied field $H_a$ (see Eq.(\ref{Us})). Therefore, $t^*$ is a decreasing function of $H_a$. These results can explain the experimental observations in Ref.~\onlinecite{Ma1} which show that: 1. $t^*$ becomes shorter and finally goes to zero with increasing $H_a$ or increasing $T$. 2. There is no concave shape in the $B(t)-t$ curves of the thin films, only convex shapes were observed. This is because the thin film samples have small surface barrier $U_{BL}$ and consequently small $U_0$, which results in the unobservable short $t^*$.

3. \textit{Effects of surface barrier on flux relaxation}.---
Eq.(\ref{Us}) shows that surface barrier affects vortex penetration. It is then natural to expect that this surface barrier also affects flux relaxation.

In flux relaxation, the external field is usually zero and the second term in Eq.(\ref{Us}) vanishes. But the first term in Eq.(\ref{Us}) still exists. The first term is caused by the surface imaging force (attractive), which helps to draw the vortices out of the superconductor and reduces the activation energy of the vortices. Early studies \cite{Ma2} have shown that in flux relaxation the time dependence of internal field is $B(t) = \sum_{l=1}^n b_l w^l(t)$, where $w(t) = U_c - kT ln\left(1 + t/\tau \right)$ and $b_l$ are constants. By further considering the surface imaging force, $w(t)$ should be replaced by
\begin{equation}
\label{FunctionW2}
w(t)= U_c - \left(\frac{\Phi_0}{4 \pi \lambda}\right)^2 K_0\left(\frac{2x}{\lambda}\right) - kT ln\left(1 + \frac{t}{\tau} \right)
\end{equation}

Eq.(\ref{FunctionW2}) indicates that the surface imaging force causes flux relaxation theory to be a local theory.

\section{Conclusion}

The general field dependent activation energy of a vortex penetration process was proposed and time dependence of internal field was derived. Although the vortex penetration is an inverse process to flux relaxation, there is no symmetry between the mathematical models of these two processes because of the surface barrier. Finally, it should be mentioned that I have ignored the effects of damping force on the vortex penetration when constructing the mathematical model. The damping force may add significant influence on the vortex penetration under large applied field where the velocity of vortices is large.

\end{document}